\documentclass[twocolumn]{aastex62}

\usepackage{amsmath}
\usepackage{natbib}

\usepackage[capitalize]{cleveref}
\usepackage{soul}
\soulregister\citet7
\soulregister\citep7
\soulregister\cref7
\soulregister\ref7

\graphicspath{{./}{figures/}}

\received{28 November 2018}
\revised{13 December 2018}
\accepted{18 December 2018}

\shorttitle{GW170817 at one year}
\shortauthors{Lamb et al.}

\begin{document}

\title{The optical afterglow of GW170817 at one year post-merger}

\correspondingauthor{GPL} \email{gpl6@leicester.ac.uk}

\author[0000-0001-5169-4143]{G.P. Lamb}
\affiliation{Department of Physics and Astronomy, University of Leicester, LE1 7RH, UK}

\author[0000-0002-3464-0642]{J.D. Lyman}
\affiliation{Department of Physics, University of Warwick, Coventry, CV4 7AL, UK}

\author{A.J. Levan}
\affiliation{Department of Physics, University of Warwick, Coventry, CV4 7AL, UK}

\author[0000-0003-3274-6336]{N.R. Tanvir}
\affiliation{Department of Physics and Astronomy, University of Leicester, LE1 7RH, UK}

\author{T. Kangas}
\affiliation{Space Telescope Science Institute, 3700 San Martin Drive, Baltimore, MD 21218, USA}

\author{A.S. Fruchter}
\affiliation{Space Telescope Science Institute, 3700 San Martin Drive, Baltimore, MD 21218, USA}

\author{B. Gompertz}
\affiliation{Department of Physics, University of Warwick, Coventry, CV4 7AL, UK}

\author[0000-0002-4571-2306]{J. Hjorth}
\affiliation{DARK, Niels Bohr Institute, University of Copenhagen, Lyngbyvej 2, Copenhagen 2100, Denmark}

\author{I. Mandel}
\affiliation{Institute for Gravitational Wave Astronomy, School of Physics and Astronomy, University of Birmingham, Birmingham, B15 2TT, UK}
\affiliation{Monash Centre for Astrophysics, School of Physics and Astronomy, Monash University, Clayton, Victoria 3800, Australia}

\author{S.R. Oates}
\affiliation{Department of Physics, University of Warwick, Coventry, CV4 7AL, UK}

\author{D. Steeghs}
\affiliation{Department of Physics, University of Warwick, Coventry, CV4 7AL, UK}

\author{K. Wiersema}
\affiliation{Department of Physics, University of Warwick, Coventry, CV4 7AL, UK}

\begin{abstract}
We present observations of the optical afterglow of GRB\,170817A, made by the {\it Hubble Space Telescope}, between February and August 2018, up to one year after the neutron star merger, GW170817. The afterglow shows a rapid decline beyond $170$~days, and confirms the jet origin for the observed outflow, in contrast to more slowly declining expectations for `failed-jet' scenarios.
We show here that the broadband (radio, optical, X-ray) afterglow is consistent with a structured outflow where an ultra-relativistic jet, with Lorentz factor $\Gamma\gtrsim100$, forms a narrow core ($\sim5^\circ$) and is surrounded by a wider angular component that extends to $\sim15^\circ$, which is itself relativistic ($\Gamma\gtrsim5$).
For a two-component model of this structure, the late-time optical decline, where $F \propto t^{-\alpha}$, is $\alpha=2.20\pm0.18$, and for a Gaussian structure the decline is $\alpha=2.45\pm0.23$.
We find the Gaussian model to be consistent with both the early $\sim10$\,days and late $\gtrsim290$ days data.
The agreement of the optical light curve with the evolution of the broadband spectral energy distribution, and its continued decline, indicates that the optical flux is arising primarily from the afterglow and not any underlying host system. 
This provides the deepest limits on any host stellar cluster, with a luminosity $\lesssim 4000 L_\odot~(M_{\rm F606W}\gtrsim-4.3)$.

\end{abstract}

\keywords{gravitational wave, relativistic processes, stars: neutron }


\section{Introduction} \label{sec:intro}

The first binary neutron star merger detected via gravitational waves (GW170817), was accompanied by a weak short-duration gamma-ray burst \citep[GRB\,170817A][]{abbott17b}, a radioactively-powered kilonova, and a long-lived afterglow \citep[e.g.][]{abbott17a}.
The steady rise of the afterglow from $\sim 10$\,days post-merger, which was traced at radio, X-ray  and optical wavelengths \citep[e.g.][]{alexander2018, davanzo2018, dobie2018, hallinan2017, lyman18, margutti2017, margutti2018, mooley2018, mooley2018a,  mooley2018b, nynka2018, piro18, resmi2018, troja2017, troja2018, vaneerten2018}, distinguished GRB\,170817A (alongside its intrinsic low-luminosity) from cosmological short-GRBs. This called into question the link between GW170817 and the progenitors of other short-GRBs.

Following a neutron-star merger, a jet, launched due to the rapid accretion of ejected matter onto a compact remnant, will propagate through the merger ejecta medium.
The interaction of the jet with the ejecta will result in a structured outflow where the wider components are the product of a cocoon of accelerated ejecta material \citep[e.g.][]{nagakura2014, murguiaberthier2017}.
The profile of this outflow depends on the mass and density of the ejected material and the initial structure of the jet. 
Simulations of jet propagation through the merger ejecta can result in outflows that have a Gaussian structure \citep{xie2018}.
This structure is responsible for driving the afterglow's evolution.
More recent simulations are beginning to reveal the structure of the jet at launch \citep{kathirgamaraju2018}.
The afterglow to GRB\,170817A is the first opportunity to convincingly probe the structure of these outflows.

For a favourably inclined gravitational-wave (GW) detected neutron-star merger, the temporal behaviour of the afterglow, viewed off the jet central axis, can probe the outflow structure and give an insight into the outflows that accompany cosmological short GRBs \citep{lamb2017,lazzati2017}.
The slow rise of the afterglow is indicative of an outflow with either an angular or radial structure.

In the angular model, the earliest afterglow observations are of the outflow components nearest to the line-of-sight.
As the outflow decelerates and expands an increasing fraction of the outflow becomes visible.
A slow rise to peak, as observed in the afterglow of GRB\,170817A, can be recreated where the angular structure of the outflow consists of a fast and energetic core (the jet, with Lorentz factor $\Gamma\gtrsim100$) and a slower, less energetic, wide component (a cocoon, $\Gamma\lesssim10$) \citep[e.g.][]{lazzati2018}.

In the radial model the outflow is wide and has a stratified velocity profile.
The fastest components ($\Gamma\sim10$) decelerate first and the resultant blast-wave is refreshed by slower components as they catch-up to the shock front.
The total energy of the blast-wave increases until the slowest component peaks; the dynamics of the final decelerating blast-wave are determined by this slowest component ($\Gamma\sim1.4 - 2.0$).
The afterglow rise following GRB\,170817A can be recreated by a wide-angled outflow with such a radial profile \citep[e.g.][]{kasliwal2017}.

At $\sim 150$\,days post-merger, the X-ray \citep{davanzo2018,margutti2018} and radio \citep{dobie2018} frequency light curves peaked and began to decline. 
Distinctive behaviour for the decline rate of the afterglow is expected depending on the dynamical and structural nature of the outflow;
a steeper decline is expected for the initially ultra-relativistic angular structured jet scenario \citep{lamb2018b, troja2018}.
A steep decline, confirming the presence of a strong jet and ruling out the wide cocoon of the radial model, was revealed by recent radio afterglow observations \citep{mooley2018b}.
The presence of an energetic jet within the outflow was additionally supported by the results from Very Long Baseline Interferometry (VLBI) of the radio source.
The super-luminal motion of the source was observed, revealing the relativistic motion of a narrow jet core launched during the merger \citep{ghirlanda2018, mooley2018a}.

In this paper we present the optical light curve of the afterglow of GRB\,170817A from \textit{Hubble Space Telescope} (\textit{HST}) imaging covering one year post-merger;
the photometry is presented in \S\ref{section2}. 
We supplement these data with radio and X-ray frequency observations to investigate the behaviour of the declining afterglow within the structured jet scenario. 
In \S\ref{section3} we fit a simple two-component jet-cocoon structure and a Gaussian structure that are both consistent with the observed data.
The Gaussian structured outflow gives a steeper decline post-peak and is more consistent with the very late-time observations at optical and radio frequencies.
The discussion and conclusions are given in \S\ref{section4} and \S\ref{section5}.

\section{Additional HST photometry}\label{section2}

\begin{figure*}
\plotone{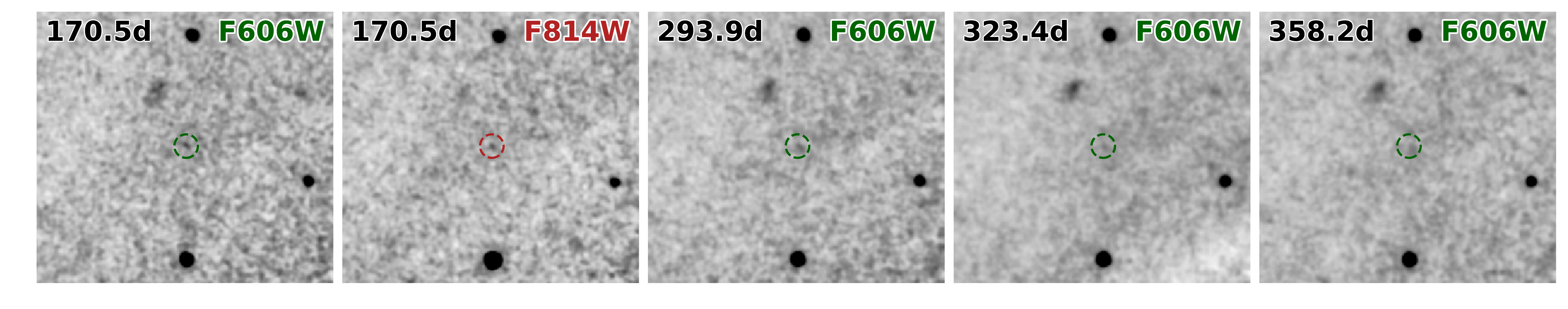}
\caption{Very late-time {\em HST} imaging of GW170817. Images are 5 arcseconds a side and the filter and rest-frame days since GW170817 are indicated in each sub-figure. An isophotal elliptical model of the smooth galaxy light has been subtracted from each figure (see text) and the images have been Gaussian-smoothed to aid the eye. Dashed circles, centred on the location of GW170817, from alignment with earlier {\em HST} epochs, are 0.25$\arcsec$ radius. North is up, East is left.}
\label{fig:hstplot}
\end{figure*}

Our \textit{HST} observations were carried out in programs GO14771 (PI: Tanvir) and GO15482 (PI: Lyman) {using WFC3 in filters F606W and F814W}. Early \textit{HST} photometry of the kilonova was presented in \citet{tanvir17} -- here we concentrate on later observations when the afterglow is dominant, extending up to one year after the merger. The first epoch of these data were presented in \citet{lyman18}, here we present four additional epochs of observations, the results of which are shown in \cref{tab:hstphot} and \cref{fig:hstplot}.
Non-detections in the near-infrared observations from December 2017 caused us to focus on optical bands for the subsequent epochs in Feburary, June, July and August 2018, corresponding to $\sim$ 171, 294, 323 and 358 days post-merger, respectively. 
Observations employed dithered exposures within visits in order to improve upon the native pixel scale using {\sc astrodrizzle} within {\sc drizzlepac}. In addition, as the July and August epochs' exposures were split across multiple visits, {\sc tweakreg} was employed to achieve accurate alignment between the visits (rms $\sim 0.10-0.15$~pixel). Further details of the reduction and analysis are presented in \citet{lyman18}.

Our photometry was performed on the drizzled images after subtracting the smooth galaxy light through isophotal ellipse fitting with the {\sc iraf} task {\sc ellipse}. A 0.08$\arcsec$ aperture was used and corrected using provided encircled energy tables\footnote{\url{http://www.stsci.edu/hst/wfc3/analysis/uvis_ee}}. Our optical light curve mimics the behaviour seen at other frequencies by peaking somewhere between our 110d and 171d epochs, before steeply declining.

For the February epoch, our photometry indicated a significant and unexpected change in the colour of the afterglow compared to our December 2017 observations. Although the F814W flux remaining almost constant, a drop (at the $\sim 3-4\sigma$ level) of $\sim 0.75$ magnitudes was seen in our F606W measurement. Inspection of the individual frames did not reveal any obvious detector artefacts. We note a near-contemporaneous measurement in F606W was made by \citet{piro18}, and, although a low significance detection at $26.4\pm0.4$ mag, this suggests no significant change of flux with respect to December 2017. Given the achromatic evolution of the synchrotron emission, we would not expect such a large colour change, particularly over neighbouring filters. Coupled with no change in the broad band evolution (i.e. radio or X-rays), we suggest these observations are most likely a statistical fluke rather than any real colour evolution in the afterglow.

Our August epoch has a marginal detection of flux and we cannot say for certain if this is entirely down to the afterglow itself, or whether some underlying surface brightness fluctuation in the galaxy light or cluster system is contributing, or at what level (indeed the source becomes visually ambiguous at this epoch, see \cref{fig:hstplot}). The measurement does however allow us to place deep constraints on any underlying host cluster, which could not be significantly brighter than the flux level we see. At a distance modulus of $\mu = 33.05\pm0.18$~mag \citep{hjorth17, cantiello18} this translates to a limit of M$_\text{F606W} = -4.3\pm0.4$ mag ($\sim 4000 L_\odot$), fainter than $\sim99\%$ of GCs found in the local group \citep[][2010 edition; see method of \citealt{lyman14}]{harris96}. 
Further, as shown later, our June epoch (294d post-merger) is almost contemporaneous with radio and X-ray measurements and our photometry for this epoch agrees well with the broadband spectral energy distribution (SED) of the afterglow. However, when subtracting the August image from our June epoch and repeating the photometry on this subtracted image, we found the resultant flux was significantly below this SED-inferred level.
This would suggest the flux is dominated by the transient, rather than any underlying fluctuation, at least at this epoch. The continued decline of the  optical flux up to the limit of our observations also indicates that the transient is the source of the flux, rather than any underlying persistent source.

\begin{deluxetable*}{lccccC}
\tablecaption{GW170817 \textit{HST} WFC3 photometry \label{tab:hstphot}}
\tablecolumns{6}
\tablehead{
\colhead{Date} & \colhead{MJD\tablenotemark{a}} & \colhead{Time since merger} & \colhead{Tot. exp. time} 
& \colhead{Filter} & \colhead{AB Mag.} \\
\colhead{} & \colhead{(d)} & \colhead{(d)} & \colhead{s} & \colhead{} & \colhead{}
}
\startdata
05 Feb 2018 & 58154.65 & 170.5 & 2400 & F814W & 26.31 \pm 0.15 \\
            & 58154.72 & 170.5 & 2400 & F606W & 27.16\tablenotemark{b} \pm 0.17 \\
10 June 2018 & 58279.27 & 293.9 & 5220 & F606W & 27.75 \pm 0.20 \\
10 July 2018 & 58309.14 &  323.4 & 14070 & F606W & 28.05 \pm 0.17 \\
14 August 2018 & 58344.23 & 358.2 & 14070 & F606W & 28.78 \pm 0.39 \\
\enddata
\tablenotetext{a}{At start of exposures.}
\tablenotetext{b}{Appears spurious considering contemporaneous data, see text.}
\tablecomments{Magnitudes have been corrected for foreground Galactic extinction following \citet{schlafly11}.}
\end{deluxetable*}

\section{Afterglow Modelling}\label{section3}

The afterglow flux of GRB\,170817A exhibits a slow rise to peak from $\sim10-150$ days as $\sim t^{0.8}$.
This behaviour is best explained by the angular structure of the outflow.
The wide angle components of a structured outflow are likely to be a cocoon of merger ejecta material that has been shocked and accelerated by the passage of an ultra-relativistic jet, where the jet is at the core of the outflow.

Post-peak, the rapid decline with an index $\alpha>1.5$, where the flux $F \propto t^{-\alpha}$, indicates that the light-curve at late-times is dominated by an initially ultra-relativistic velocity jet or core of the outflow \citep{lamb2018b}.
A rapid post-peak decline has been confirmed by X-ray and radio observations \citep{nynka2018, alexander2018, vaneerten2018, mooley2018b} and here via optical observations with {\it HST}.

We determine light-curves for two angular structured outflow models that give good fits to the data.
Motivated by the VLBI observations of a super-luminal core with a half-opening-angle $\theta_c\lesssim 5^\circ$ and observed at an inclination from the outflow central axis $\iota\sim 20^\circ$ \citep{mooley2018a}, we limit the range for these two parameters in our model fits to $0.6^\circ\leq\theta_c\leq6^\circ$ and the inclination $17^\circ\leq\iota\leq23^\circ$.
With these tight constraints we consider the models:
\begin{itemize}
\item Model (A) - a simple two-component structure consisting of a narrow uniform $<\theta_c$, energetic and ultra-relativistic $\Gamma\gtrsim100$ core surrounded by a wide, relativistic cocoon $\Gamma=5$ with $10\%$ of the core energy per steradian over angles $\theta_c - \theta_j$, where $\theta_j=15^\circ$ and is the edge of the outflow.\footnote{A energetic cocoon that can account for the afterglow light-curve $t\lesssim80$\,days and with an initial Lorentz factor of $\Gamma<5$ will dominate the late-time decline resulting in a decay index $\alpha\lesssim2.0$ \cite{lamb2018b}. Such a decline is ruled out by the data}
\item Model (B) - a Gaussian structure where the energy per steradian is $\propto e^{-\theta^2/\theta_c^2}$ and the Lorentz factor is $\propto e^{-\theta^2/2\theta_c^2}$ (and condition $\Gamma>1$) within $\theta_j$.
\end{itemize}
Model (A) is a simple structure based on \cite{lazzati2017a,lazzati2017} and \cite{lamb2017} where the Lorentz factor of the cocoon is $\Gamma<10$;
the fixed parameters ensure the cocoon is energetic enough to contribute at early times and reduces the number of parameters in the fit.
Model (B) was used originally in relation to the GRB\,170817A afterglow by \cite{resmi2018} and \cite{lamb2018a}.
We limit the opening angle of the outflow to $\sim15^\circ$;
for Model (A), a much wider cocoon would require a more complex structure, and for Model (B), the low energetics of wider components would contribute insignificantly to the light-curve.

The afterglow flux from each model is calculated using an updated version of the structured outflow method described in \citep{lamb2017, lyman18}.
The dynamics for the expanding blast-wave follow the analytic solution in \cite{peer2012}, includes sideways expansion at the sound speed\footnote{Sideways expansion of the outflow is required at late-times as the decline is $\alpha\gtrsim1.6$, the limit expected for a jetted outflow with these parameters. Using a more realistic expansion \cite[e.g.][]{vaneerten2012B} will have only a small effect on the fitted parameters}, and a more accurate synchrotron flux estimation \citep[see][]{lamb2018b}).

We use a Markov Chain Monte Carlo (MCMC) {\sc emcee} to determine the best parameter fits for each model using the flux at 3 and 6 GHz \citep{dobie2018, margutti2018, mooley2018, mooley2018b}, the {\it HST} optical data points \citep[this study;][]{lyman18, piro18}, and {\it Chandra} X-ray flux at 1 keV \citep{margutti2017, troja2017, troja2018}.
For each model we fit eight parameters,
\begin{equation*}
\Phi = \left[~E_{\rm iso,c},~ \Gamma_{0, c},~ \theta_c,~ \iota,~ \varepsilon_B,~ \varepsilon_e,~ n_0,~ p~\right],
\end{equation*}
where $E_{\rm iso,c}$ is the isotropic equivalent energy of the central core point, $\Gamma_{0, c}$ is the bulk Lorentz factor pre-deceleration of the central core point, $\varepsilon_B$ and $\varepsilon_e$ are the microphysical parameters, $n_0$ is the ambient medium particle density, and $p$ the accelerated electron power-law index.\footnote{Parameters $E_{\rm iso,c},~\varepsilon_B,~ \varepsilon_e,~{\rm and}~ n_0$ have logarithmic flat priors, while $\Gamma_{0,c},~ \theta_c,~ \iota,~{\rm and}~p$ have flat priors.
We use 40 walkers, 2000 burn in steps and 15000 steps per model}

The parameter constraints for each model are shown in \cref{tab:emcee} where the uncertainties represent the 16$^{\rm th}$ and 84$^{\rm th}$ percentiles.
We see the expected correlations and degeneracies within the parameter distributions i.e. $\varepsilon_B$ with $n$ and both $\varepsilon_B$ and $n$ with the core energy.
Model (A) favours an inclination towards the upper limit within our range and $\varepsilon_e$ is pushed against the upper bound;
whereas for Model (B) we see a positive correlation between core angle and inclination and find that the core energy, Lorentz factor and jet core angle favour the upper half of the parameter range. 

\begin{figure}
\plotone{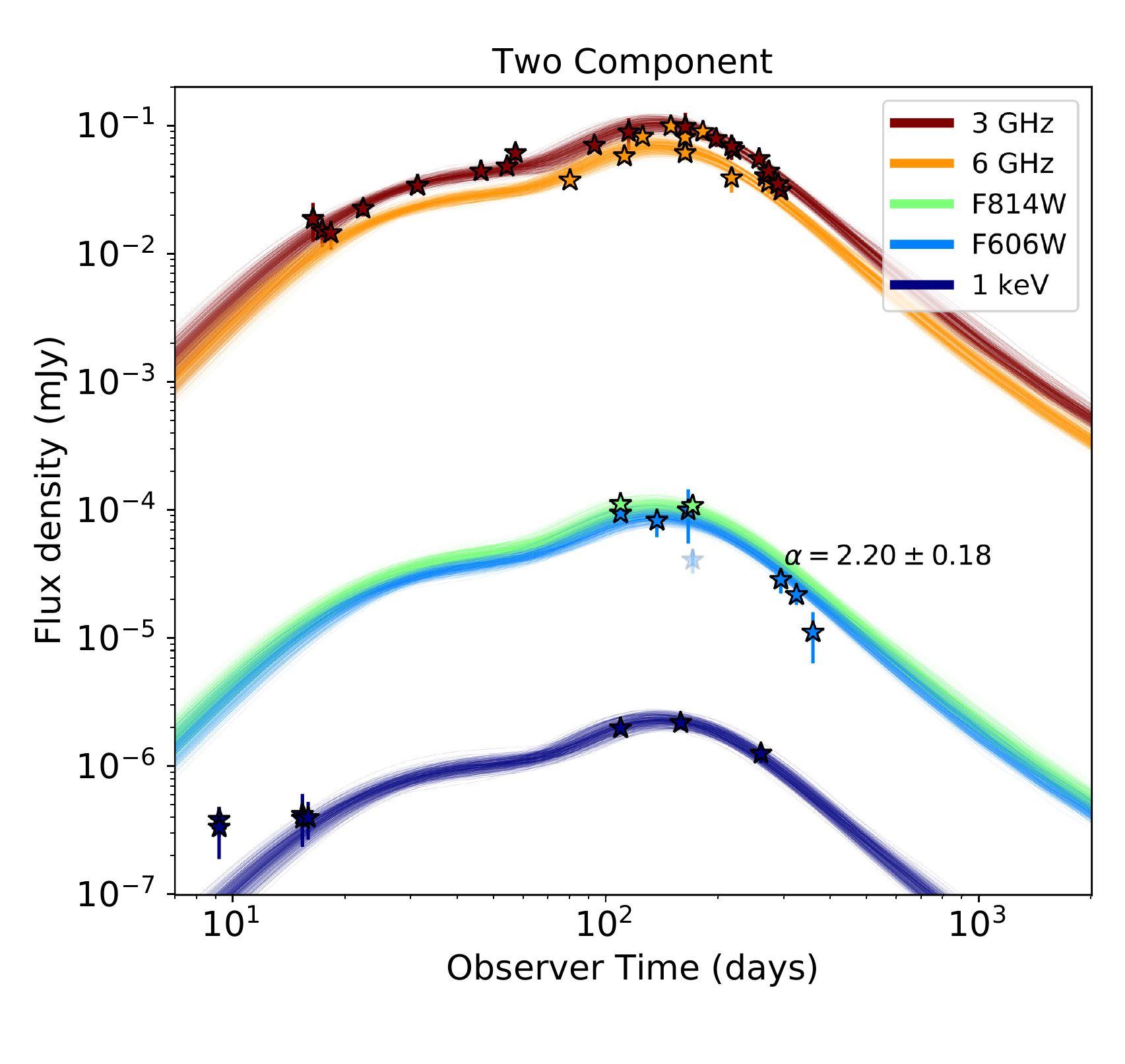}
\caption{Model light-curves for 400 randomly selected parameter sets from an MCMC for Model (A). Stars represent the data points and error bars are 1$\sigma$ (error bars may be hidden by the markers). Light-curves at 3 GHz, 6 GHz, $3.8\times10^{14}$\,Hz (F814W), $5.1\times10^{14}$\,Hz (F606W), and 1 keV are shown. The decline index $\alpha$ between 260-300 days is annotated. The faint point at $\sim170$\,days shows the anomalous F606W point discussed in \S \ref{section2}.}
\label{lcfitA}
\end{figure}

\begin{figure*}
\plotone{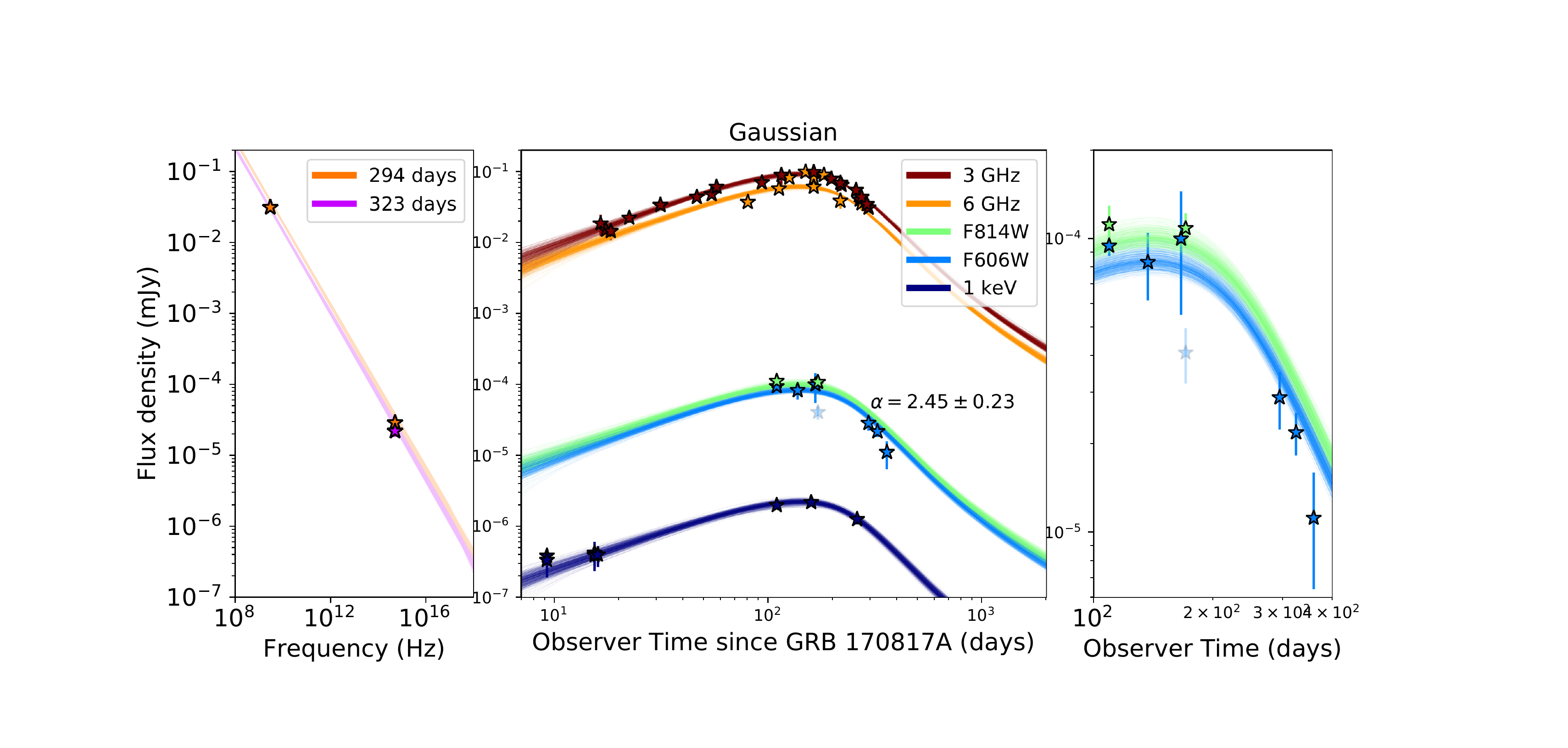}
\caption{[Left panel] 400 SEDs at 294 days (red line) and 323 days (purple line). Stars show data for each epoch, error bars are smaller than the marker size.
[Middle panel] Model light-curves for 400 randomly selected parameter sets from an MCMC for Model (B). Data as \cref{lcfitA}. The decline index $\alpha$ between 260-300 days is annotated. [Right panel] Zoom of the optical data and light-curves between 100 and 400 days post GRB\,170817A/GW170817. }
\label{fig:lcfit}
\end{figure*}

For each model we show 400 light-curves using randomly selected parameters drawn from the sample;
light-curves for Model (A) are shown in \cref{lcfitA} and the light-curves for Model (B) in \cref{fig:lcfit}.
The light-curve at 3 GHz, $3.8\times10^{14}$\,Hz (F814W), $5.1\times10^{14}$\,Hz (F606W) and 1\,keV are shown with data as stars.
The temporal decline $\alpha$, calculated between $260-300$\,days at optical frequencies, is $\alpha=2.20\pm0.18$ for Model (A) and $\alpha=2.45\pm0.23$ for Model (B). 
The optical data and the Model (B) light-curve, and the model SED at 294 and 323 days are highlighted in \cref{fig:lcfit}.

\begin{deluxetable*}{lcccC}[t!]
\tablecaption{Inferred afterglow parameters \label{tab:emcee}}
\tablecolumns{4}
\tablehead{
\colhead{Parameter} & \colhead{Prior range} & \colhead{Model (A)} & \colhead{Model (B)} & \colhead{unit}
}
\startdata
$\log_{10}E_{\rm iso,c}$ & 50 -- 53 & $52.0^{+0.6}_{-0.9}$ & $52.4^{+0.4}_{-0.5}$ & $\log_{10} {\rm erg}$ \\
$\Gamma_{0,c}$ & 10 -- 1000 & $88^{+40}_{-28}$ & $666^{+231}_{-272}$ & \\
$\theta_c$ & 0.01 -- 0.1 & $0.07^{+0.01}_{-0.01}$ & $0.09^{+0.01}_{-0.01}$ & ${\rm rad}$ \\
$\iota$ & 0.3 -- 0.4 & $0.36^{+0.03}_{-0.03}$ & $0.34^{+0.02}_{-0.02}$ & ${\rm rad}$ \\
$\log_{10}\varepsilon_B$ & -4 -- -0.5 & $-2.4^{+1.4}_{-0.9}$ & $-2.1^{+0.8}_{-1.0}$ & \\
$\log_{10}\varepsilon_e$ & -4 -- -0.5 & $-1.3^{+0.6}_{-0.7}$ & $-1.4^{+0.5}_{-0.6}$ & \\
$\log_{10}n_0$ & -5 -- 0 & $-3.3^{+0.6}_{-1.0}$ & $-4.1^{+0.5}_{-0.5}$ & $\log_{10}{\rm cm}^{-3}$ \\
$p$ & 2.01 -- 2.25 & $2.17^{+0.01}_{-0.01}$ & $2.16^{+0.01}_{-0.01}$ &
\enddata
\tablecomments{Subscript `c' indicates the jet central point. Parameter values are the median from {\sc emcee} distributions, uncertainties represent the 16$^{\rm th}$ and 84$^{\rm th}$ percentiles.}
\end{deluxetable*}

\section{Discussion}\label{section4}
We have presented optical observations made by {\it HST} of the afterglow to GRB\,170817A between February and August 2018, see \cref{tab:hstphot}.
using this data we confirm the rapid decline of the afterglow indicative of an initially ultra-relativistic jet viewed off-axis.
Combining the optical data with radio wavelength observations at 3 and 6\,GHz, and X-ray frequency data at 1\,keV, we use {\sc emcee} to fit two outflow structure models, see \cref{lcfitA} and \ref{fig:lcfit}. 

The post-peak decline seen here at optical frequencies, and at radio frequencies by \cite{mooley2018b}, is rapid;
the decline index $\alpha\gtrsim2.1$. 
An $\alpha\sim p$ is expected for an on-axis observed afterglow following the jet-break, a steeper decline is expected for an observer outside of the initial jet half-opening angle \citep[e.g.][]{vaneerten2012} and the latest data points indicate a very rapid decline at late times.  
Such a rapid decline requires an initially ultra-relativistic and collimated outflow -- a jet -- and rules out the possibility here of a wide-angled mildly relativistic outflow \citep{lamb2018b}.

The two outflow models used to fit the afterglow data show differing late-time declines.
The two-component Model\,(A) shows a shallower decline at $\alpha\sim2.2$, and the Gaussian-structured outflow Model\,(B) shows a steeper decline at $\alpha\sim 2.5$, between 260-300 days post-merger.
{If the decline is shown to steepen at later epochs then the expansion description or the jet core structure should be reconsidered.}

The origin of the gamma-ray emission in GRB\,170817A, is debated. 
A faint GRB would be an expectation for an off-axis observation.
However, the spectral peak energy for the prompt emission and the lack of an early afterglow challenges the simple off-axis model \citep[e.g.][]{ioka2018, lamb2018a, matsumoto2018, nakar2018}.
The leading explanations for the prompt origin include:
a short GRB seen off-axis but considering more complex emission models \citep[e.g.][]{eichler2018,zhang2018}; 
a GRB scattered by cocoon material \citep{kisaka2018};
and a burst of gamma-rays as a result of a cocoon shock breakout \citep{gottlieb2018}.
It is beyond the scope of this work to determine the origin of the prompt emission, although the steep sides of the core in both models (A) and (B) are consistent with the description \cite{kisaka2018} required for scattered prompt emission.

We have tested only angular structure models;
this is supported by the success of Model\,(A) at reproducing the early afterglow data.
Where the Lorentz-factor of the cocoon is $<5$ the late-time afterglow decline is shallower than $\alpha\lesssim2$ due to the contribution of the cocoon.
This supports the need for a relativistic outflow from core to edge.
We note that the Gaussian structure can account for all of the data from 10 days, whereas the two-component model fails to recreate the first X-ray frequency data points, see \cref{lcfitA}.

Both models have a peak $\sim140-160$ days, and predict a rapid decline $\alpha\gtrsim2.0$.
The transition to a Newtonian blast-wave is seen more prominently in the Gaussian model at $\sim700$ days post-merger, although this is below the detection threshold at all frequencies.
The counter-jet will contribute to the light-curve beyond the range of the figures, at  $\sim 10^4$\,days.


\section{Conclusions}\label{section5}

{\it HST} observations of the afterglow of GRB\,170817A, taken from 171 days to one year from merger GW170817, show it to be rapidly declining in flux. We find the declining optical flux is most consistent with arising from the afterglow, matching the behaviour seen at other frequencies, and thus can be used to place the most stringent constraints on any underlying globular cluster system, which must be $M_\textrm{F606W} \gtrsim -4.3\pm0.4$~mag.

We have modelled the afterglow using both a two-component jet model, consisting of a narrow highly-relativistic core and wider-angle component, and a Gaussian structured outflow. Both scenarios are able to broadly recreate
the steep decline post-peak; such a steep decline requires an initially collimated, highly-relativistic outflow and confirms a successful jet was launched in GRB\,170817A, in agreement with other lines of evidence \citep{ghirlanda2018, mooley2018a}. We find most consistency with the Gaussian outflow to describe our very-late time photometry (although uncertainties in the measurements are large for these epochs).

\acknowledgments{Based on observations made with the NASA/ESA Hubble Space Telescope, obtained from the data archive at the Space Telescope Science Institute. STScI is operated by the Association of Universities for Research in Astronomy, Inc. under NASA contract NAS 5-26555. These observations are associated with programs GO\,15482 (Lyman), GO\,14771 (Tanvir). 
GPL is supported by STFC grant ST/N000757/1 and thanks Om Salafia for useful discussions. 
JDL acknowledges support from STFC grant ST/P000495/1.
AJL acknowledges that this project received funding from the European Research Council (ERC) under the European Union's Horizon 2020 research and innovation programme (grant agreement 725246).
JH acknowledges support by a VILLUM FONDEN Investigator grant (project number 16599).
SRO acknowledges support of the Leverhulme Trust Early Career Fellowship. 
We thank the anonymous referee for their comments.}

%

\vspace{5mm}
\facilities{HST(WFC3)}


\software{astropy \citep{astropy},
		  DrizzlePac \cite{drizzlepac},
		  emcee \citep{emcee},
          SExtractor \citep{sextractor}
          }




\bibliographystyle{aasjournal}





\end{document}